\documentclass[aps,onecolumn,secnumarabic,nobalancelastpage,amsmath,amssymb,
nofootinbib]{revtex4}

\usepackage{graphics}      
\usepackage{graphicx}      
\usepackage{longtable}     
\usepackage{url}           
\usepackage{bm}            

\usepackage{amsmath}
\usepackage[brazil]{babel}
\usepackage[utf8]{inputenc}

\usepackage{color}

\newtheorem{lemma}{Lemma}

\begin{document}
\title{The (restricted) Inomata-McKinley spinor representation and the underlying topology}
\author         {D. Beghetto}
\email          {dbeghetto@feg.unesp.br}
\author         {J. M. Hoff da Silva}
\email          {hoff@feg.unesp.br}
\affiliation    {Universidade Estadual Paulista - UNESP, Departamento de F\'{i}sica e Qu\'{i}mica, Guaratinguet\'{a}, SP, Brazil.}
\date{\today}

\begin{abstract}
The so called Inomata--McKinley spinors are a particular solution of the non-linear Heisenberg equation. In fact, free linear massive (or mass-less) Dirac fields are well known to be represented as a combination of Inomata--McKinley spinors. More recently, a subclass of Inomata--McKinley spinors were used to describe neutrino physics. In this paper we show that Dirac spinors undergoing this restricted Inomata--McKinley decomposition are necessarily of the first type, according to the Lounesto classification. Moreover, we also show that this type one subclass spinors has not an exotic counterpart. Finally, implications of these results are discussed, regarding the understanding of the spacetime background topology.
\end{abstract}

\maketitle

\section{introduction}

The very idea of Inomata-McKinley decomposition \cite{Ino} was devoted to understand neutrino physics in the early days, in the context of Wheeler geometrodynamics \cite{Whee}. More recently, this decomposition was employed to construct Dirac linear fermions via non-linear special Heisenberg spinors\footnote{By non-linear spinors, we mean spinors obeying non-linear dynamics.} \cite{Nove}. In fact, the so called Inomata-McKinley spinors are a subclass of non-linear Heisenberg spinors. In Ref. \cite{Nove}, after showing that Dirac fields may be decomposed in terms of Inomata-McKinley spinors, six disjoint topological sectors, in which the decomposed fields may reside, are constructed. This result turns out to be physically appealing since, after analyzing the corresponding helicities, each topological sector embraces spinorial fields describing neutrino (and anti-neutrino) states. When compared to the original procedure, the decomposition presented in Ref. \cite{Nove} can be faced as a particularization, whose importance rests upon its physical implications. This particularization (the main topic to be investigated here) is what we call restricted Inomata-McKinley decomposition.   

Given the physical relevant aspect of Dirac spinorial fields written in terms of Inomata-McKinley spinors, it would be useful to understand which type of Dirac spinors may the used for such. As a matter of fact, there are (physically and geometrically) different Dirac spinors allowed in the four-dimensional spacetime. In a true work of categorization, Lounesto worked out a physical classification of spinor fields. This classification, differently from others, is particularly important due to the bilinear covariants \cite{Lou}. According to this classification, there are exactly six different spinorial field classes. One of the goals of this work is to show that all Dirac spinors undergoing the restricted Inomata-McKinley (RIM) representation, used to describe neutrino physics, are necessarily type one, in the aforementioned classification. Roughly speaking, these type of spinors have more interactions possibilities, allowing self-interaction terms and couplings scrutinizing parity symmetry. 

Moreover, the decomposition itself is strongly dependent of the triviality of the background topology. This is reflected in the fact that the currents (the usual one and the chiral one) must be irrotational \cite{Nove}. Therefore, in the case of a non-trivial topology, all the aforementioned representation of usual Dirac fields fails off. On the other hand,  within nontrivial topology context there exists an exotic spinorial structure, whose elements are exotic counterparts of usual spinors \cite{Exo}. Briefly stating, the exotic spinorial structure is due to the necessary different patches of local coverings and, therefore, usual and exotic spinors coexist and the difference between them may be charged to the dynamical equation. In fact, all the topological non triviality is traduced in a new term on the Dirac operator which can be understood, since it has the same net effect, as a vectorial coupling. In this sense, it would be plausible that new currents taking into account the topological term could open a crevice, leading to the representation of exotic spinors in terms of Inomata-McKinley ones. As shall be seen, we show that exotic spinors cannot undergo the RIM decomposition. This result is more appealing than it may sound. In fact, by the reason previously exposed, it is a genuine problem to separate out the usual from the exotic spinors. Our result suggests that as far as neutrino physics is well described by the RIM procedure, this very physical system can serve as a criteria to set (at least locally) the underlying topology. We shall elaborate on that in the final remarks. The two main results obtained were chosen to be exposed as lemmata, for the sake of clarity.  

This paper is organized as follows: a brief review of Lounesto's classification and RIM spinors are presented in Section 2. We show in Section 3 a strong constraint in representing a Dirac spinor in terms of RIM spinors. Section 4 is devoted to the study of exotic spinors and the impossibility of decompose them in terms of RIM spinors. In the last section we conclude.

\section{Elementary review}

For bookkeeping purposes, we shall describe the basic introductory elements, pointing out the main necessary aspects to reach our conclusions. 

\subsection{The Lounesto classification of spinors}

The program elaborated by Lounesto, categorizing spinors as elements belonging to six different sectors of the spinorial space, was reviewed and studied in a vast literature \cite{Lou} (for a modern view point, see \cite{onthe}). Let $\Psi$ be a spinor, whose bilinear covariants read
\begin{itemize}
 \item[1)] $A = \Psi^\dagger \gamma_0 \Psi,$
 \item[2)] $\textbf{J} = J_\mu \theta^\mu = \Psi^\dagger \gamma_0 \gamma_\mu \Psi \theta^\mu,$
 \item[3)] $\textbf{S} = S_{\mu \nu} \theta^{\mu \nu} = \frac{1}{2} \Psi^\dagger \gamma_0 i \gamma_{\mu \nu} \Psi \theta^\mu \wedge \theta^\nu,$
 \item[4)] $\textbf{K} = K_\mu \theta^\mu = \Psi^\dagger \gamma_0 i \gamma_{0123} \gamma_\mu \Psi \theta^\mu, $
 \item[5)] $B = - \Psi^\dagger \gamma_0 \gamma_{0123} \Psi,$
\end{itemize}
where $\{x^\mu\}$ is a set of global spacetime coordinates, in a given inertial frame $\textbf{e}_\mu = \frac{\partial}{\partial x^\mu}$, and the set $\{ \theta^\mu \}$ represents the dual basis of $\{ \textbf{e}_\mu \}$. These bilinear covariants are not completely independent. In fact, defining a multivector structure, $Z = A + \textbf{J} + i\textbf{S} - i \gamma_{0123}\textbf{K} +  \gamma_{0123}B$, it is possible to see that the following identities hold:
\begin{eqnarray}
Z^2 = 4 \sigma Z;\\
Z \gamma_\mu Z = 4 J_\mu Z;\\
Z i \gamma_{\mu \nu} Z = 4S_{\mu \nu} Z;\\
Z \gamma_{0123} Z = -4 \omega Z;\\
Z i \gamma_{0123} \gamma_\mu Z = 4 K_\mu Z.
\end{eqnarray}
 
The appreciation of the constraints above allowed the classification of $\Psi$ into the classes (for which $\textbf{J}$ is always non-zero):
 \begin{itemize}
  \item[1)] $A \neq 0; \;\;\; B \neq 0.$
  \item[2)] $A \neq 0; \;\;\; B = 0.$
  \item[3)] $A = 0; \;\;\; B \neq 0.$
  \item[4)] $A = 0 = B; \;\;\; \textbf{K} \neq 0; \;\;\; \textbf{S} \neq 0.$
  \item[5)] $A = 0 = B; \;\;\; \textbf{K} = 0; \;\;\; \textbf{S} \neq 0.$
  \item[6)] $A = 0 = B; \;\;\; \textbf{K} \neq 0; \;\;\; \textbf{S} = 0.$
 \end{itemize}
For the classes 1, 2 and 3, it holds $\textbf{K}, \textbf{S} \neq 0$, and the spinors bellowing to these classes are called regular spinors. The classes 4, 5 and 6 consist of the so-called singular spinors. Usual spinors describing fermions in field theory have place in classes 1, 2, and 3. As we mentioned, in Section 3 we demonstrate that all Dirac spinors undergoing an Inomata--McKinley decomposition, in trying to describe neutrinos, belong to class 1 exclusively. As it is possible to envisage from the scheme above, this class allows for the richer coupling arrangements possible. This may help neutrino model builders in study many physical interactions.   
 

\subsection{Dirac linear fermions and RIM spinors}

The Dirac equation of motion is well known to be linear with respect to the spinor fields. Its nonlinear counterpart, the so-called Heisenberg equation, is given by  
\begin{eqnarray}\label{heisenberg}
[i\gamma^{\mu}\partial_{\mu}-2s(A+iB\gamma^{5})]\Psi^{H}=0,
\end{eqnarray}

\noindent with $A\equiv\overline{\Psi}^{H}{\Psi}^{H}$ and $B\equiv i\overline{\Psi}^{H}\gamma^{5}{\Psi}^{H}$ being the usual  bilinear covariants associated to $\Psi^H$. The constant $s$ has dimension $(\text{lenght})^2$. The Heisenberg equation (\ref{heisenberg}) can be properly obtained by varying the action constructed from the lagrangian 
\begin{equation}
L = \frac{i}{2}\bar{\psi}\gamma^\mu\partial_\mu\psi-\frac{i}{2}\partial\bar{\psi}\gamma^\mu\psi-sJ_\mu J^\mu,
\end{equation} with respect to the spinor field \cite{novelodela,Hei}. In Ref. \cite{Heialemao}, the nonlinear spinor equation was deeply investigated under discrete symmetries. Without assuming any particular symmetry of the spinor fields, it is shown that the dynamical equation itself is invariant under $C$, $P$, and $T$ symmetries. It is also shown that the theory is invariant under scale transformation, which is in a good agreement with the perspective of using massless fields (see Section \ref{sectionFinalRemarks}).

The RIM solution of the Heisenberg equation (\ref{heisenberg}) is a particular class of solutions given by
\begin{eqnarray}\label{inomata}
\partial_{\mu}\Psi=(aJ_{\mu}+bK_{\mu}\gamma^{5})\Psi,
\end{eqnarray}

\noindent with $a,b \in \mathbb{C}$. A spinor $\Psi$ satisfying the condition (\ref{inomata}) shall be called a RIM spinor. This is because in the original decomposition the first term in the right-hand side of (\ref{inomata}) is given by $K^\lambda\gamma_\lambda \gamma_\mu \gamma^5$ and the mapping between this last term and $J_\mu$ is not so direct, being in fact given by a regular nontrivial matrix operator, say $G$. For instance, starting from $J_{\mu}$ it is possible to enlarge the decomposition by means of 
\begin{equation}
G=\frac{1}{2J^2}J^\mu K^\nu[\gamma_\nu,\gamma_\mu]\gamma^5. 
\end{equation} However, we shall keep our analysis in terms of RIM spinors due to its physical appealing. It is possible to prove that every RIM spinor is a solution of the Heisenberg equation (\ref{heisenberg}) with $2s=i(a-b)$. The integrability condition of (\ref{inomata}) requires the constraint $\text{Re}(a) = \text{Re}(b)$.

Moreover, it is necessary a regular behavior to the covariant currents, i. e., $J^{\mu}$ and $K^{\mu}$ must be irrotational. Hence, denoting the norm $J^{2}=J_{\mu}J^{\mu}$, one defines $J_{\mu}=\partial_{\mu}S$, with $S=\frac{1}{(a+\overline{a})}\ln\sqrt{J^{2}}$ being a scalar. In an analog fashion, $K_{\mu}=\partial_{\mu}R$, with $R=\frac{1}{(b-\overline{b})}\ln\bigg(\frac{A-iB}{\sqrt{J^{2}}}\bigg)$. Yet, using the notation $J \equiv \sqrt{J^2}$, according to Ref. \cite{Nove} it is possible to represent a Dirac spinor $\Psi^D$ by
\begin{eqnarray}
\Psi^{D}=\exp{\left[\frac{iM}{(a+\overline{a})J}\right]}J^{2\sigma}\bigg(\sqrt{\frac{J}{A-iB}}\Psi^{H}_{L}+\sqrt{\frac{A-iB}{J}}\Psi^{H}_{R}\bigg), \label{esse}
\end{eqnarray}

\noindent where $\Psi^H$ is a RIM spinor (which has left-hand $\Psi_L$ and right-hand $\Psi_R$ components), $M$ is the mass parameter coming from the Dirac equation, $J^{2\sigma} = \exp{\left[ \left(2is - \frac{b-\bar{b}}{2}\right)S \right]}$, and $\sigma \equiv -\frac{i\;\text{Im}(a)}{4\;\text{Re}(a)}$. Therefore, Dirac spinors can be represented as a combination of RIM spinors, which satisfy the Heisenberg non-linear equation (\ref{heisenberg}). Interestingly enough, such a procedure reveals important to describe neutrino physics \cite{Ino,Nove}.


\section{Constraints on representing Dirac fields in terms of RIM spinors}

In order to see to which class the spinor (\ref{esse}) belongs, it is necessary to construct its associated bilinear covariants in terms of $\Psi^H$. First of all, notice that
\begin{eqnarray}
 \Psi^D = \alpha J^{2\sigma} \left(\sqrt{\frac{J}{A-iB}} \Psi^H_L + \sqrt{\frac{A-iB}{J}} \Psi^H_R \right),
\end{eqnarray}

\noindent with $\alpha \equiv \exp{\left[i \frac{M}{2\text{Re}(a) J} \right]}$. Also, one can have the representation $\Psi^H_L = \frac{1}{2}(\mathbb{I} + \gamma^5) \Psi^H$ and $\Psi^H_R = \frac{1}{2}(\mathbb{I} - \gamma^5) \Psi^H$. Thus, denoting $\beta \equiv \sqrt{\frac{J}{A-iB}}$, we have
\begin{eqnarray}
 \Psi^D = \alpha J^{2\sigma} \left[\beta (\mathbb{I} + \gamma^5) + \beta^{-1} (\mathbb{I} - \gamma^5) \right] \Psi^H.
\end{eqnarray} Now, recalling that 
\begin{equation}
\sqrt{z} = \sqrt{|z|}\; \frac{z + |z|}{|(z + |z|)|},
\end{equation} i. e., 
\begin{eqnarray}
\text{Re}(\sqrt{z}) = \sqrt{|z|}\; \frac{\text{Re}(z) + |z|}{|(z + |z|)|},\;\;\text{Im}(\sqrt{z}) = \sqrt{|z|}\; \frac{\text{Im}(z)}{|(z + |z|)|},
\end{eqnarray} for $z \in \mathbb{C}$, it is possible to see that
\begin{eqnarray}
 \left( \Psi^D \right)^{\dagger} =\left( \Psi^H \right)^{\dagger} \left\{ \frac{2}{\sqrt{2J(J+A)}} \left[(J+A) + B\gamma^5 \right] \right\} \left( J^{2\sigma} \right)^{-1} \alpha^{-1}. \label{algumacoisa}
\end{eqnarray}

With (\ref{algumacoisa}) at hands, one can construct the bilinear covariants associated to the Dirac spinors. After some calculations they read 
\begin{eqnarray}\label{bilinearAD}
 A_D & = & T_{(ABJ)} {\bar{\Psi}^H} \left[(A+J-B)(\mathbb{I} + \gamma^5) + \frac{(A+J+B)(A-iB)}{J}(\mathbb{I} - \gamma^5) \right] \Psi^H,\\\label{bilinearBD}
 B_D & = & i T_{(ABJ)} {\bar{\Psi}^H} \left[(A+J-B)(\mathbb{I} + \gamma^5) - \frac{(A+J+B)(A-iB)}{J}(\mathbb{I} - \gamma^5) \right] \Psi^H,\\\label{bilinearJD}
 J_D^{\mu} & = & T_{(ABJ)} {\bar{\Psi}^H} \gamma^{\mu} \left[(A+J+B)(\mathbb{I} + \gamma^5) + \frac{(A+J-B)(A-iB)}{J}(\mathbb{I} - \gamma^5) \right] \Psi^H,\\\label{bilinearKD}
 K_D^{\mu} & = & -i T_{(ABJ)} {\bar{\Psi}^H} \gamma^{\mu} \left[(A+J+B)(\mathbb{I} + \gamma^5) - \frac{(A+J-B)(A-iB)}{J}(\mathbb{I} - \gamma^5) \right] \Psi^H,
\end{eqnarray}

\noindent with the scalar $T_{(ABJ)} \equiv \sqrt{\frac{2}{(J+A)(A-iB)}} \in \mathbb{C}$.

Now, notice that by construction (see Section 2) it is impossible to have both $A = 0$ and $B = 0$ simultaneously, i. e., RIM spinors are indeed regular spinors. Therefore, it cannot be the case of $A - iB = 0$ to occur. Still, obviously $J = \sqrt{A^2 + B^2} \neq 0$. Bearing these constraints in mind, let us show that $A_D, B_D, J_D^{\mu}$ and $K_D^{\mu}$ are all necessarily non-vanishing. 

Firstly, suppose that $A + J - B = 0$ in Eqs. (\ref{bilinearAD}-\ref{bilinearKD}). In this way we have  $A + J + B = 2B$. In order to have $A + J + B = 0$, we have to set $B = 0$, which leads to the condition $A + J = 0$. However, if $B = 0$, then it turns out that, necessarily, $A \neq 0$, and since $J \neq 0$, it yields $A + J \neq 0$ (see $T_{(ABJ)}$), evincing a contradiction. Therefore, $A + J + B \neq 0$, and none of the Dirac bilinears vanish. On the other hand, by a quite similar reasoning, if we suppose that $A + J + B = 0$ in Eqs. (\ref{bilinearAD}-\ref{bilinearKD}), it is straightforward to see that $A + J - B = -2B$. Then, in order to have $A + J - B = 0$, we have to set $B = 0$, which leads to the condition $A + J = 0$. Again, this procedure makes explicit a contradiction. Hence, none of the Dirac bilinears vanish.

Therefore, we have just proved the following:
\begin{lemma}
 Every Dirac spinor, acting in an usual spacetime (there is, with trivial topology), written in terms of RIM spinors, belongs to the class 1 in the Lounesto classification.
\end{lemma}


\section{Exotic spinors structures and RIM spinors}

It is consensual that half-integers representations of the Poincar\'e group are subtle. This subtlety reveals itself, among innumerable other issues, in treating fermions whose dynamics is taken in a spacetime endowed with nontrivial topology \cite{Exo}. Let us recall the main aspects of exotic spinors\footnote{For a complete account on the existence of additional spinorial structures see \cite{nois}.}. A given manifold, when multiply connected, may have many spinor bundles, wich are split into equivalence classes, the so-called spin structures. The set of spin structures is labeled by the cohomology group $H^1(\pi_1(M),\mathbb{Z}_2)$ elements, where $M$ is the base manifold (the spacetime here) and $\pi_1$ stands for the first homotopy group.   

Naturally, the manifestation of the nontrivial topology is regarded to the spin connection, thus affecting the derivative operator. After all, the net effect of nontrivial topology in which concerns the spinor dynamics may be inputed to a $1-$form extra `coupling' \cite{Exo}. The difference between the dynamics of the usual Dirac spinors $\Psi$ and of the exotic spinors $\tilde{\Psi}$, then, comes from the fact that the connection related to $\tilde{\Psi}$ must feel the non-trivial topology. Hence, the exotic dynamical equation can be written as
 \begin{eqnarray}\label{exotic1}
  [i(\gamma^\mu \partial_\mu + \gamma^\mu \partial_\mu \theta) - m \mathbb{I}] \tilde{\Psi} = 0.
 \end{eqnarray} In Eq. (\ref{exotic1}), the $0-$form $\theta$ is directly linked to the nontrivial topology and, therefore, by setting $\theta = 0$ the Dirac equation is recovered. 

One of the attempts to construct a new condition on the currents, taking into account the topology exoticness, may be obtained by inspecting the derivative term which leads to an invariant $\tilde{J}^\mu\equiv\bar{\tilde{\Psi}}\gamma^{\mu}\tilde{\Psi}$. Firstly, Eq. (\ref{exotic1}) has its conjugate, which can be written as
 \begin{eqnarray}\label{exoticConj1}
  -i\partial_\mu \tilde{\Psi}^{\dagger} (\gamma^\mu)^{\dagger} - i \tilde{\Psi}^{\dagger} \partial_\mu \theta (\gamma^\mu)^{\dagger} = m \tilde{\Psi}.
 \end{eqnarray} Multiplying Eq. (\ref{exoticConj1}) by $\gamma^0$ from the right side, it leads to
\begin{eqnarray}\label{exoticConj2}
  \partial_\mu \bar{\tilde{\Psi}} \gamma^\mu = i m \bar{\tilde{\Psi}} - \bar{\tilde{\Psi}} \partial_\mu \theta \gamma^\mu.
 \end{eqnarray} Now, using Eqs. (\ref{exoticConj1}) and (\ref{exoticConj2}) it follows straightforwardly that
  \begin{equation}\label{exoticJ}
 \partial_\mu \tilde{J}^\mu = -2 \partial_\mu \theta \bar{\tilde{\Psi}} \gamma^\mu \tilde{\Psi} \Rightarrow (\partial_\mu  + 2 \partial_\mu \theta) \tilde{J}^\mu = 0.
 \end{equation} Therefore, since $\theta$ is an arbitrary scalar function, one can redefine $2\theta \mapsto\theta$ and write the new operator which leaves $\tilde{J}^\mu$ invariant as 
 \begin{equation}\label{exoticNabla}
  \tilde{\nabla}_\mu  = \partial_\mu + \partial_\mu \theta.
 \end{equation} Notice that, in fact, the functional form of Eq. (\ref{exoticNabla}) could be advised from the appreciation of the derivative operator present in Eq. (\ref{exotic1}).
 
It will be proved that one cannot have exotic spinors written in terms of RIM spinors. Firstly, the analog condition of Eq. (\ref{inomata}) for the case of exotic spinors is therefore the following
\begin{eqnarray}\label{inomataExotico}
\partial_{\mu}\Psi=(aJ_{\mu}+bK_{\mu}\gamma^{5} - \partial_\mu \theta)\Psi
\end{eqnarray} and the irrotational currents conditions (non-longer valid in a multiply connected topology) are to be replaced by  
\begin{eqnarray}
 \tilde{J}_\mu = \partial_\mu S + S \partial_\mu \theta,\label{irrotational1} \\
\tilde{K}_\mu = \partial_\mu R + R \partial_\mu \theta,\label{irrotational2}
\end{eqnarray}

\noindent with $S$ and $R$ scalar quantities. Thus, one wants to know the explicit form of these scalars. It is readily simple to use the condition (\ref{inomataExotico}) to obtain
\begin{equation}\partial_\mu \tilde{J}_\nu = (\bar{a}\tilde{J}_{\mu}+\bar{b}\tilde{K}_{\mu}\gamma^{5} + \partial_\mu \theta)\bar{\tilde{\Psi}} \gamma_\nu \tilde{\Psi} + \bar{\tilde{\Psi}} \gamma_\nu (a\tilde{J}_{\mu}+b\tilde{K}_{\mu}\gamma^{5} + \partial_\mu \theta)\tilde{\Psi},\end{equation}

\noindent leading to 
\begin{eqnarray}\label{exotico1}
\partial_\mu \tilde{J}_\nu = (a + \bar{a})\tilde{J}_\mu \tilde{J}_\nu + (b + \bar{b})\tilde{K}_\mu \tilde{K}_\nu + 2\partial_\mu \theta \tilde{J}_\nu.
\end{eqnarray}

Multiplying Eq. (\ref{exotico1}) by $\tilde{J}^\nu$ from the right side, it turns into
\begin{eqnarray}
\frac{1}{2}\frac{\partial_\mu \tilde{J}^2}{\tilde{J}^2(a+\bar{a})} = \tilde{J}_\mu + \frac{2}{(a+\bar{a})}\partial_\mu \theta.\label{abo}
\end{eqnarray} Nevertheless, $\tilde{J}_\mu = \partial_\mu S + S \partial_\mu \theta$ and then, substituting this result into Eq. (\ref{abo}), and noticing that $\frac{\partial_\mu \tilde{J}^2}{\tilde{J}^2} = \partial_{\mu} \ln{(\tilde{J}^2)}$, one obtains 
\begin{eqnarray}\label{exotico2}
 S \partial_{\mu}\theta = \partial_{\mu} \left[-S + \frac{1}{2(a+\bar{a})} \ln{(\tilde{J}^2)} - \frac{2\theta}{(a+\bar{a})} \right].
\end{eqnarray} Finally, note that the quantity between brackets on the right hand side of Eq. (\ref{exotico2}) is a scalar. In this vein, let $H$ be the scalar defined as
\begin{eqnarray}\label{H}
 H = -S + \frac{1}{2(a+\bar{a})} \ln{(\tilde{J}^2)} - \frac{2\theta}{(a+\bar{a})}.
\end{eqnarray} One is certainly allowed to write
\begin{eqnarray}
 S \partial_{\mu}\theta = \partial_{\mu} H \Rightarrow J_{\mu} = \partial_{\mu} (S+H),
\end{eqnarray}

\noindent which means that $\tilde{J}_{\mu}$ is irrotational, making explicit a conflict with equation (\ref{irrotational1}).

On the other hand, with $\tilde{A} \equiv \bar{\tilde{\Psi}} \tilde{\Psi}$ and $\tilde{B} \equiv -\bar{\tilde{\Psi}} \gamma_{0123} \tilde{\Psi}$, as it can be readily verified, using equation (\ref{inomataExotico}) one can obtain 
$$\partial_\mu \tilde{A} = (\partial_\mu \bar{\tilde{\Psi}})\tilde{\Psi} + \bar{\tilde{\Psi}}(\partial_\mu \tilde{\Psi}),$$ leading to
\begin{eqnarray}\label{A-exotico}
 \partial_\mu \tilde{A} = (a+\bar{a})\tilde{J}_\mu \tilde{A} + i(b-\bar{b})\tilde{K}_\mu \tilde{B} - 2(\partial_\mu \theta)\tilde{A}.
\end{eqnarray}
Analogously, 
$$\partial_\mu \tilde{B} = i(\partial_\mu \bar{\tilde{\Psi}}) \gamma_5 \tilde{\Psi} + i \bar{\tilde{\Psi}} \gamma_5 (\partial_\mu \tilde{\Psi})$$ and one is able to write 
\begin{eqnarray}\label{B-exotico}
 \partial_\mu \tilde{B} = (a+\bar{a})\tilde{J}_\mu \tilde{B} + i(b-\bar{b})\tilde{K}_\mu \tilde{A} - 2(\partial_\mu \theta)\tilde{B}.
\end{eqnarray} Contracting Eq. (\ref{A-exotico}) by $\tilde{K}_\mu$ and Eq. (\ref{B-exotico}) by $(-i\tilde{K}_\mu)$ from the right and summing up the result, we have 
\begin{eqnarray}
 \tilde{K}_\mu = \frac{1}{(b+\bar{b})} \partial_\mu \left[ \ln{\left(\frac{|\tilde{A} - i\tilde{B}|}{\tilde{J}}\right)} \right],
\end{eqnarray}

\noindent which means that $\tilde{K}_{\mu}$ is also irrotational, leading to a conflict with Eq. (\ref{irrotational2}).

Therefore, Eqs. (\ref{irrotational1}) and (\ref{irrotational2}) are no longer valid in this scenario. Notice that, if they hold, exotic spinors would behave as usual ones. Thus, it is proved the following Lemma valid on a base manifold with non-trivial topology:

\begin{lemma}
The representation of exotic spinors by a combination of RIM spinors is completely obstructed.
\end{lemma}

We shall elaborate on this result in the next section. 

\section{Final remarks}\label{sectionFinalRemarks}

In this paper we have shown two lemmata related to the physics of neutrinos, when described by restricted Inomata--Mckinley spinors. First, we have found that from among the several classes of spinors, Dirac fermions decomposed in terms of RIM spinors are always of type one according to the so-called Lounesto classification. This particular class have all the bilinear covariants necessarily non vanishing and, therefore, the richer multivectorial structure. From the physical point of view, it means that all possible couplings are likely to be studied. This is, in fact, an important remark. Indeed, it allows for the study of bigger set of perturbatively renormalizable couplings in neutrino physics. 

Moreover, we also investigated the behavior of RIM fields in a non-trivial base manifold, where exotic spinors are expected. In fact, it is not trivial to separate out usual from exotic spinors, when both are allowed in a physical system. There are some exceptions \cite{excecao,nois}, but they are not very common. Here, we have a typical physical system without any exotic counterpart. In fact, it was demonstrated that it is not possible to implement the restricted Inomata--Mckinley decomposition in such a context. The point to be stressed here is that it is an important negative result, since by describing neutrino physics, in an acceptable ground, via the RIM decomposition we are {\it ab initio} necessarily fixing the base topology as trivial. It means that neutrino physics may serve as a tool to probe the spacetime topology, at least locally. In this context, neutrino physics act as a complement of Elko exotic spinors, whose additional couplings come exclusively from nontrivial topology \cite{nois}. In fact, the whole investigation performed in Ref. \cite{Nove} points to the existence of six disjoint types of spinors which can be grouped together into two major helicity sectors (composed by three types, each). These sectors can be connected by means of an unitary linear transformation indicating the possibility of describing neutrinos oscillation even in the massless neutrinos case \cite{Nove}. Bearing in mind the exhaustive result presented in Lemma 2, the eventual observation of massless neutrinos indicates -- necessarily -- the triviality of the underlying topology. In other words, a non-trivial topology is incompatible with oscillation of massless neutrinos. 

\section*{Acknowledgments}
The authors are grateful to Prof. Roldao da Rocha for useful conversation. JMHS thanks to CNPq (304629/2015-4; 445385/2014-6) for partial financial support.

\end{document}